# TOM: An Open-Source Tongue Segmentation Method with Multi-Teacher Distillation and Task-Specific Data Augmentation

Jiacheng Xie[a,b], Ziyang Zhang[c], Biplab Poudel[a,b], Congyu Guo[a,b], Yang Yu[a,b], Guanghui An[d], Xiaoting Tang[e], Lening Zhao[f], Chunhui Xu[b,g], Dong Xu[a,b,g]*

[a] Department of Electrical Engineering and Computer Science, University of Missouri, Columbia, MO, USA;
[b] Christopher S. Bond Life Sciences Center, University of Missouri, Columbia, MO, USA;
[c] Department of Computer Science, McCormick School of Engineering, Northwestern University, Chicago, IL, USA;
[d] School of Acupuncture and Tuina, Shanghai University of Traditional Chinese Medicine, Shanghai, China;
[e] Community Health Service Center Shanghai Pudong New Area, Shanghai, China;
[f] Yingcai Honors College, University of Electronic Science and Technology of China, Chengdu, China;
[g] Institute for Data Science and Informatics, University of Missouri, Columbia, MO, USA
*Corresponding author at: Department of Electrical Engineering and Computer Science, University of Missouri, Columbia, MO, 65211 USA.

## Abstract

Tongue imaging serves as a valuable diagnostic tool, particularly in Traditional Chinese Medicine (TCM). The quality of tongue surface segmentation significantly affects the accuracy of tongue image classification and subsequent diagnosis in intelligent tongue diagnosis systems. However, existing research on tongue image segmentation faces notable limitations, and there is a lack of robust and user-friendly segmentation tools. This paper proposes a **to**ngue image segmentation **m**odel (TOM) based on multi-teacher knowledge distillation. By incorporating a novel diffusion-based data augmentation method, we enhanced the generalization ability of the segmentation model while reducing its parameter size. Notably, after reducing the parameter count by 96.6% compared to the teacher models, the student model still achieves an impressive segmentation performance of 95.22% mIoU. Furthermore, we packaged and deployed the trained model as both an online and offline segmentation tool (available at https://itongue.cn/), allowing TCM practitioners and researchers to use it without any programming experience. We also present a case study on TCM constitution classification using segmented tongue patches. Experimental results demonstrate that training with tongue patches yields higher classification performance and better interpretability than original tongue images. To our knowledge, this is the first open-source and freely available tongue image segmentation tool.

# 1. Introduction

Tongue diagnosis, a fundamental diagnostic method in traditional Chinese medicine (TCM) (Tang et al., 2008), has also gained increasing attention in modern medical research. Medical tongue diagnosis makes inferences and diagnoses of some diseases by observing the manifestations on the patient's tongue. In clinical practice, physicians make inferences and analyses of the patient's health condition based on the color, shape, texture, and wetness of the tongue. In terms of tongue color, strawberry tongue is associated with scarlet fever and staphylococcal sepsis (Adya et al., 2018); map tongue is related to papillitis of the tongue (Núñez Amin Dick et al., 2021); and black hairy tongue is one of the black uncommon fungal infections (Ren et al., 2020). The cracks in the tongue are a typical textural abnormality that is strongly associated with Melkersson-Rosenthal syndrome (Ozgursoy et al., 2009), Down's syndrome (Avraham et al., 1988) and diabetes mellitus (Farman, 1976). Tongue shapes such as teeth marks provide a wealth of disease-related information (*The Association of Tongue Scalloping With Obstructive Sleep Apnea and Related Sleep Pathology - Todd M. Weiss, Strahil Atanasov, Karen H. Calhoun, 2005*, n.d.; Tomooka et al., 2017). Tongue diagnosis has been widely used in medical diagnosis as a non-invasive, intuitive, and rapid diagnostic method.

Currently, most achievements in intelligent tongue diagnosis are based on tongue image segmentation, which assists in the clinical evaluation of a subject's health condition. Accurate tongue segmentation is crucial for subsequent clinical decisions and diagnostic tasks. There are two primary scenarios for tongue image segmentation: (1) professional tongue diagnostic instruments and (2) other specialized acquisition devices. These devices typically capture tongue images from a fixed distance in a standardized environment, and this collection method is significantly limited by spatial and equipment conditions. The other scenario is open-environment tongue image acquisition, which has become popular with the widespread use of mobile acquisition devices such as smartphones. Due to the varying resolutions of different mobile devices, differing lighting conditions, and the presence of numerous extraneous objects in the images, these factors pose significant challenges to the accurate segmentation of the tongue. The primary challenges in tongue segmentation primarily stem from two aspects. Firstly, the tongue's color closely resembles surrounding areas, such as the lips and face, complicating the segmentation task. Additionally, the rich morphological variations of the tongue, including abnormal shapes and diverse surface textures, further challenge the segmentation process. Secondly, in open environments, the noise introduced by background elements and complex lighting conditions during image acquisition also poses significant difficulties.

Tongue image segmentation represents a critical task in the digitalization of tongue diagnosis. With the advancement of computer vision technology, methodologies for tongue segmentation have evolved progressively from traditional graphics-based methods to machine learning, especially deep learning, and, more recently, large-scale pre-trained models. In the early 21st century, tongue segmentation primarily relied on traditional image processing techniques, which involved segmentation based on features such as tongue color, edges, and shape. Common methods included thresholding (Al-amri et al., 2010), edge detection (H. Zhang et al., 2006), region growing (Tremeau & Borel, 1997), and active contour models (Zuo et al., 2004). However, these traditional techniques were susceptible to variations in illumination and typically required substantial manual parameter adjustments, resulting in relatively weak generalization capabilities. Around 2010, with developments in pattern recognition and machine learning, feature extraction-based machine learning approaches, such as support vector machines (SVM) (Bertelli et al., 2011), K-means clustering (Ibragimov et al., 2015), and random forests (Mahapatra, 2014), were introduced for tongue segmentation tasks. Nevertheless, the performance of these methods depended significantly on the quality of manually designed features, which restricted their adaptability and generalizability.

Since 2015, breakthroughs in deep learning methodologies within the computer vision domain have led to convolutional neural network (CNN)-based tongue segmentation methods becoming mainstream. These included segmentation frameworks derived from CNN architectures such as VGG (Shi et al., 2021), ResNet (Lin et al., 2018), end-to-end segmentation approaches such as U-Net and its improved variants (Azad et al., 2024), and self-attention-based segmentation models exemplified by Vision Transformer and Swin Transformer (He et al., 2022). While these deep learning methods obviated manual feature engineering, they incurred substantial training costs and necessitated considerable computational resources. In 2023, Meta AI introduced the Segment Anything Model (SAM) (Kirillov et al., 2023), a large-scale pre-trained model capable of zero-shot segmentation. This model enables automatic segmentation of target regions through interactive prompts on large foundational pre-trained models. However, the original SAM exhibits suboptimal performance when directly applied to specific medical imaging tasks, thus necessitating further fine-tuning to enhance its suitability and efficacy in medical applications.

To address the above challenges, we propose a **to**ngue segmentation optimization **m**odel (TOM) based on multi-teacher distillation, as shown in **Fig. 1**. Initially, our method employs prompt boxes generated by an object detection model as inputs. It performs lightweight fine-tuning on the original SAM model, enhancing its adaptability to the tongue segmentation task. Subsequently, we introduce a novel hybrid multi-teacher distillation strategy, utilizing the fine-tuned SAM model, UNet, and DeepLabV3 as teacher models. This strategy significantly reduces the parameter scale of the student model while maintaining segmentation performance comparable to the teacher models. Furthermore, during training, we use diffusion models to generate synthetic tongue images for data augmentation. Compared to traditional augmentation techniques, this method further enhances model generalization and effectively mitigates overfitting risks. Finally, we apply this optimized model to TCM constitution classification tasks based on tongue images. Experimental results demonstrate that the classification models trained on segmented tongue images significantly outperform those trained directly on original tongue data in terms of accuracy and interpretability. Following deployment, TCM practitioners and researchers can achieve streamlined, end-to-end tongue segmentation tasks with a single-click solution. The main contributions of this paper are as follows:

(1) We proposed a novel tongue image segmentation method based on multi-teacher distillation, which achieves satisfactory segmentation performance on a broader range of open-source and private datasets. Furthermore, we release two models: a larger model that enables precise local segmentation and a smaller model that supports web-based online segmentation. To the best of our knowledge, this is the first tongue image segmentation tool that is freely available for doctors without requiring any coding effort.

(2) Compared to traditional image data augmentation methods, we propose a diffusion-based data augmentation approach designed explicitly for tongue segmentation tasks. This method effectively enhances the segmentation model's adaptability across different scenarios and mitigates overfitting issues caused by data scarcity.

(3) We present a case study on the application of tongue segmentation, demonstrating that in the constitution classification task of tongue diagnosis, using segmented tongue data for classification can further improve model performance compared to using raw tongue images.

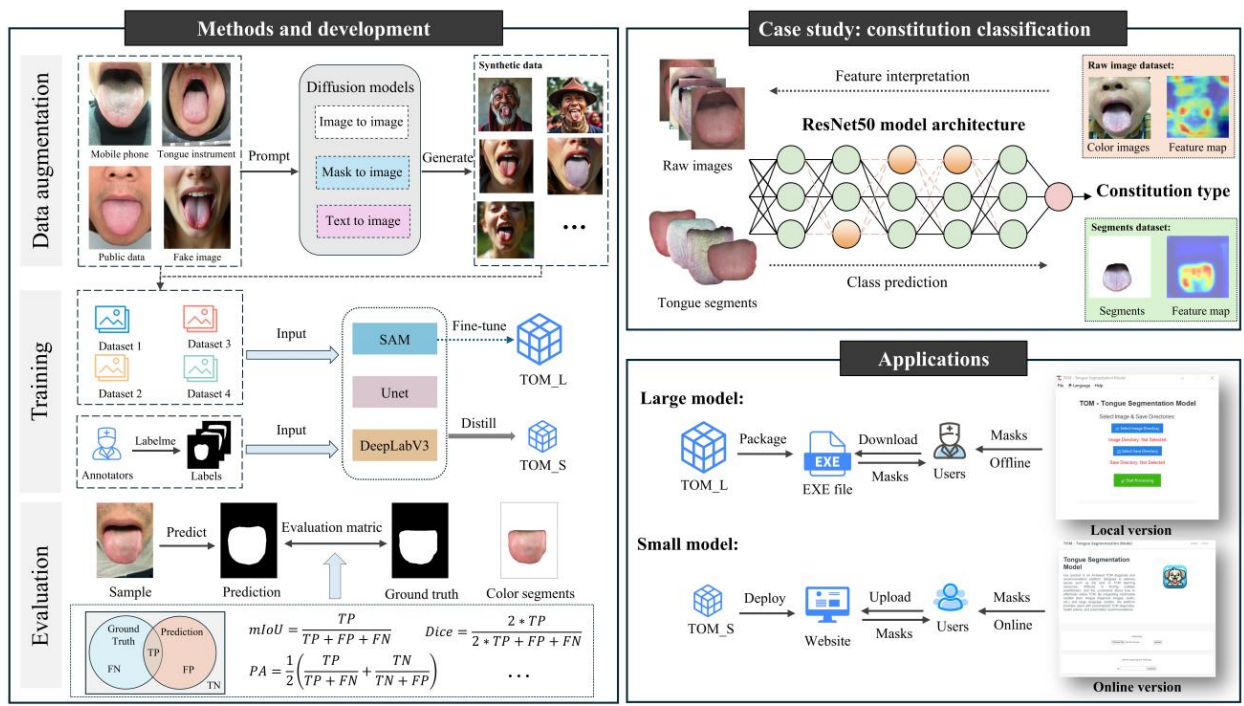

**Fig. 1. Workflow of TOM.** We proposed a novel data augmentation method based on diffusion models to enhance the diversity of tongue image data and reduce the risk of model overfitting. The augmented data was combined with the original dataset and annotated with segmentation masks by TCM experts. We fine-tuned a SAM-based teacher model, TOM_L, using the annotated dataset. Then, through multi-teacher distillation with UNet and DeepLabV3, we obtained a lightweight student model, TOM_S. They were deployed in a local application and an online platform, respectively. Additionally, a case study on TCM constitution classification showed that models trained on segmented tongue patches achieved significantly better accuracy and interpretability than those trained on raw tongue images.

## 2. Related work

Recent advances such as SAM and Knowledge Distillation (KD) offer promising solutions for segmentation and model optimization. This section reviews their core principles, applications in segmentation, and inherent limitations to contextualize our proposed approach.

### 2.1. Segment anything model

Built on a Vision Transformer (ViT) (Dosovitskiy et al., 2021), SAM includes an image encoder, prompt encoder, and mask decoder, and is trained on 11 million images and 1 billion masks. Its strength lies in generalizing to unseen objects without retraining, making it promising for medical imaging where data is limited and diverse. However, in specialized tasks like tongue segmentation, domain-specific fine-tuning is often required due to subtle color differences, morphological variability, and complex backgrounds.

Several studies have explored SAM's potential in medical imaging tasks such as CT, MRI, ultrasound, and X-rays. MedSAM (Ma et al., 2024) improves segmentation accuracy by fine-tuning SAM on large-scale medical datasets but faces challenges with low-contrast boundaries and high computational costs. SAMed (K. Zhang & Liu, 2023) leverages low-rank adaptation (LoRA) for efficient parameter tuning in tasks like polyp segmentation. Other adaptations, including SkinSAM (Hu et al., 2023) for dermatology and SAM_Path (*SAM-Path: A Segment Anything Model for Semantic Segmentation in Digital Pathology | SpringerLink*, n.d.) for histopathology, further demonstrate SAM's adaptability to specific medical domains.

Building on this progress, SAM has also been adapted for tongue image segmentation. TongueSAM (Cao et al., 2023) combines an object detection–based prompt generator with SAM to enable automated zero-shot segmentation, achieving high accuracy but remaining sensitive to prompt quality and computationally intensive. Tongue-LiteSAM (Tan et al., 2025) offers a lightweight alternative to reduce resource demands while maintaining reasonable performance, though it struggles with generalization across diverse tongue morphologies. MEAT-SAM (Zhong et al., 2025) aims to balance accuracy and efficiency on edge devices but faces limitations in detail preservation under low-contrast conditions. These adaptations underscore SAM's promise for TCM-related tongue segmentation while also highlighting trade-offs between accuracy, efficiency, and adaptability.

## 2.2. Knowledge Distillation

Knowledge distillation (KD) (Hinton et al., 2015) is a widely adopted model compression technique that transfers knowledge from a large teacher model to a smaller student model, maintaining comparable performance while significantly reducing the computational cost. By learning from the teacher's outputs rather than ground truth labels, KD enables efficient deployment in resource-constrained environments. Knowledge can be transferred through various forms, including response-based (output logits), feature-based (intermediate representations), and relation-based (structural dependencies).

In medical imaging, KD has proven effective in addressing the demands of deep learning models. Qin et al. (Qin et al., 2021) proposed a segmentation framework that distills semantic information from a pretrained teacher to a lightweight student model, demonstrating high efficiency. The integration of KD with SAM has also gained attention. Julka et al. (Julka & Granitzer, 2024) applied KD to transfer SAM's prompt-based segmentation knowledge to a domain-specific decoder for geological mapping, while Chen et al. (Chen & Bai, 2023) utilized KD to adapt SAM for thermal infrared segmentation. In the medical domain, Huang et al. (Huang et al., 2025) introduced KnowSAM, combining SAM with a co-teaching strategy for semi-supervised segmentation, enabling effective multi-view knowledge exchange. Other approaches such as MobileSAM (C. Zhang et al., 2023) and TinySAM (Shu et al., 2025) further illustrate how KD enhances SAM's deployability and efficiency across diverse scenarios.

## 3. Proposed Approach

### 3.1. Diffusion-based data augmentation method

Data augmentation enhances model robustness and generalization by generating diverse samples through transformations of original data. Classic methods (Buslaev et al., 2020), such as geometric manipulations, color adjustments, noise addition, and occlusions, have shown some benefits in tongue image analysis. However, they offer limited diversity, risk distorting anatomical structures, and cannot modify high-level semantic attributes. Inappropriate augmentations may reduce clinical utility by obscuring diagnostic features or introducing artifacts (J. Li et al., 2023).

Earlier studies employed GANs to synthesize tongue images for data-limited scenarios, but suffered from instability and mode collapse due to low diversity. In contrast, diffusion models generate data by learning a denoising process, offering superior image quality and semantic variability. Unlike traditional augmentations, they can produce medically realistic variations in coating thickness, color, cracks, and teeth marks. Despite their success in medical image generation, applications of diffusion models for tongue image augmentation remain scarce. To address this, we propose a novel augmentation approach designed for tongue segmentation tasks, as illustrated in **Fig. 2**, comprising three key components.

### 3.1.1. Image-to-image data augmentation

The training and inference processes of diffusion models consist of two main phases: forward diffusion and reverse diffusion. In the forward diffusion process, given a real tongue image $x_0$, Gaussian noise is added progressively at each timestep $t$ (from 1 to $T$), generating a sequence of noisy images $x_t$:

$$q(x_t|x_{t-1}) = \mathcal{N}(x_t; \sqrt{1-\beta_t}x_{t-1}, \beta_t I) \tag{1}$$

where $\beta_t$ is a predefined noise schedule controlling the amount of noise added at each step. When $t \to T$, $x_t$ approaches pure Gaussian noises.

In the reverse diffusion process, A neural network, typically a UNet, learns the denoising process to gradually reconstruct the tongue image from pure noise:

$$p_\theta(x_{t-1}|x_t) = \mathcal{N}(x_{t-1}; \mu_\theta(x_t, t), \Sigma_\theta(x_t, t)) \tag{2}$$

where $\mu_\theta(x_t, t)$ predicts the denoised mean, and $\Sigma_\theta(x_t, t)$ estimates the variance.

The experiment in this study is conducted using a publicly available tongue image dataset and follows a structured process. First, data preprocessing is performed to standardize the tongue images, including resizing and color normalization, ensuring compatibility with the model input. Next, during model training, a UNet-based denoising network is employed, incorporating time embeddings to encode timestep information. A fixed linear or cosine noise schedule is used, and the optimization objective is to minimize the denoising error, represented as:

$$\mathbb{E}_{x_0, t, \epsilon}[||\epsilon - \epsilon_\theta(x_t, t)||^2] \tag{3}$$

where $\epsilon_\theta$ is the noise predicted by the neural network, and $\epsilon$ is the actual noise added. Finally, in the image generation and evaluation phase, new tongue images are synthesized through reverse diffusion, starting from pure noise. The quality of the generated images is assessed using structural similarity index (SSIM) and Frechet inception distance (FID) to measure their similarity to real tongue images. Meanwhile, to further enhance the quality of data augmentation, we introduce a manual screening mechanism to review the generated tongue images. This process ensures that only high-quality images, which accurately reflect real tongue characteristics in terms of color, texture, and structure, are selected. The screened high-quality images are then incorporated into the augmented dataset to improve the quality of training data for subsequent model training, thereby enhancing the model's generalization ability in tongue image analysis tasks.

### 3.1.2. Mask to image

Generating new images through local image replacement and synthesis also constitutes a viable approach for data augmentation. Several early studies have employed this technique within AI-based face manipulation methods, including DeepFake (Seow et al., 2022) and FaceSwap (Korshunova et al., 2017). Such methods enable the replacement of actors' facial features, application to diverse roles, and adjustment of facial expressions, fundamentally relying on GAN or Diffusion Models for facial feature segmentation, alignment, and blending. Similar methodologies have been utilized in AI-powered clothing try-on tasks; for instance, Hsiao et al. (Hsiao et al., 2019) introduced a diffusion model-based approach for virtual garment fitting. In the medical imaging domain, some studies have applied these techniques for data augmentation purposes. Wang et al. (J. Wang et al., 2025) successfully enhanced the detection accuracy of HER2-positive breast cancer by employing generative models to synthesize diverse, high-quality breast cancer MRI images. Similarly, Wu et al. (J. Wu et al., 2024) proposed a segmentation method leveraging diffusion models, achieving superior performance across over 20 medical image segmentation tasks.

In this section, we introduce a tongue image style-transfer and generation approach based on diffusion models. As illustrated in **Fig. 2**, our method utilizes diffusion models to generate backgrounds corresponding to tongue image masks, thereby achieving image style transfer. Specifically, the proposed framework leverages stable diffusion 1.5 (SD15) or stable diffusion XL (SDXL) as the foundational generation model. BrushNet (*BrushNet*, n.d.) is incorporated to enhance image quality and modulate the style, color, and texture through positive and negative clip embeddings, ensuring visual consistency with targeted tongue characteristics. Additionally, we utilize ComfyUI (Xue et al., 2024), a powerful graphical user interface for Stable Diffusion, facilitating visualization of the image generation pipeline via node-based workflows. The complete workflow for the data augmentation process, along with our model implementation, has been made publicly available.

### 3.1.3. Text to image

The third component of our data augmentation method relies exclusively on textual prompts for generating synthetic tongue image data. To this end, we have constructed a prompt pool containing a series of carefully crafted, high-quality prompts designed to ensure that the generated tongue images closely resemble real human tongue photographs in terms of color, morphology, and surface characteristics. Moreover, to enhance the diversity of generated images and reduce stylistic redundancy, we developed a prompt optimizer that inserts random-length ASCII sequences into randomly selected positions within each prompt, thereby guaranteeing the uniqueness of each prompt used during image synthesis. This approach to prompt optimization has been validated in several related studies (*Design Guidelines for Prompt Engineering Text-to-Image Generative Models | Proceedings of the 2022 CHI Conference on Human Factors in Computing Systems*, n.d.; Jiang et al., 2024) and has shown to be effective for altering the prompt structure in each generation cycle and identifying optimal prompt formulations. Subsequently, these optimized prompts are utilized across multiple AI-based image generation platforms, including Midjourney, DALL-E 3, Adobe Firefly, Imagen 3, and NightCafe, to produce synthetic tongue images exhibiting diverse visual styles and characteristics.

Following the image-generation process, the generated images undergo rigorous assessment and selection by professional TCM experts rather than directly incorporating all synthetic images into tongue segmentation tasks. Only images closely resembling real-world photographic conditions are selected and integrated into the expanded dataset for further augmentation.

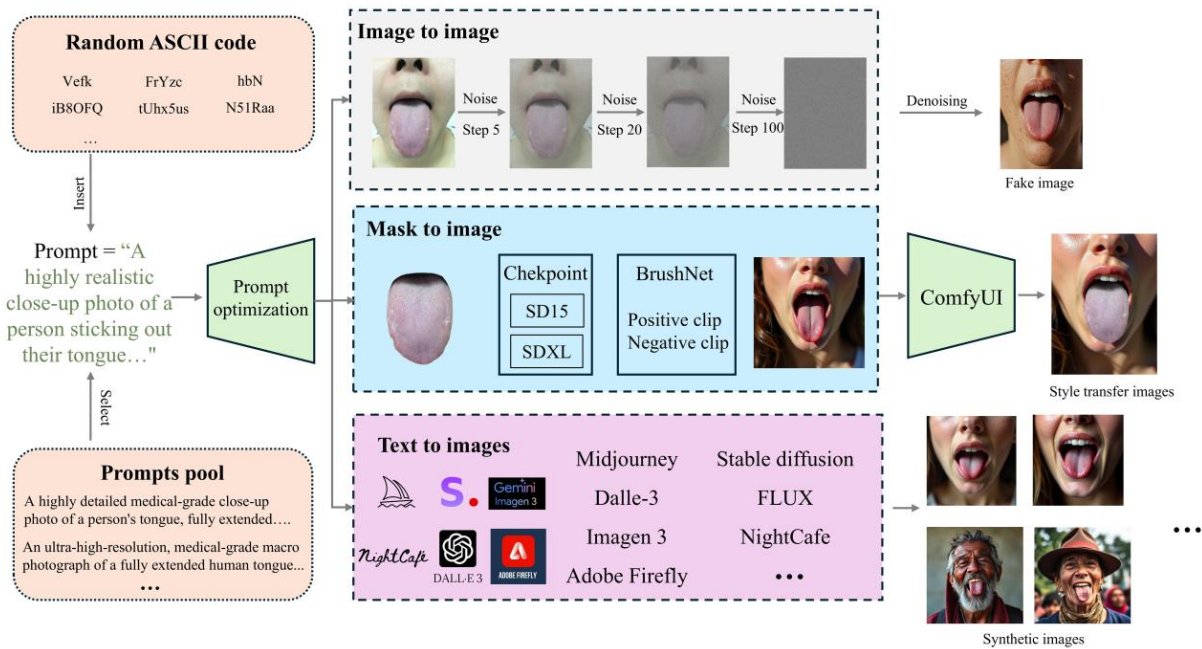

**Fig. 2. A diffusion model-based tongue image augmentation framework leveraging prompt optimization and multi-condition generation strategies.** This pipeline enhances tongue image datasets by synthesizing diverse, high-quality samples through diffusion models. It begins with optimized prompts derived from either random ASCII codes or a curated prompt pool describing medically relevant tongue images. These prompts guide three conditional image generation paths: (1) *Image-to-Image*, where existing tongue images undergo progressive noising and denoising to produce realistic variations; (2) *Mask-to-Image*, which uses segmented tongue masks and checkpoints (e.g., SD15, SDXL) for fine-grained control; and (3) *Text-to-Image*, which produces synthetic tongue images via models such as DALL·E 3, Midjourney, and Stable Diffusion. Outputs are further refined with ComfyUI for style transfer, enriching visual diversity while preserving anatomical realism—ultimately supporting robust tongue image classification, segmentation, and diagnostic tasks.

## 3.2 SAM-based segmentation method

In this subsection, we propose a SAM-based segmentation method, where the trained model, TOM_L, also serves as one of the teacher models in the subsequent multi-teacher distillation task. As illustrated in **Fig. 3**, the proposed method comprises two components: a prompt generation module and a segmentation module. The prompt generation module is designed to localize the tongue region within the image, providing bounding boxes and points as input prompts for further processing. In the segmentation module, the original SAM is fine-tuned to better accommodate the specific characteristics of tongue image segmentation.

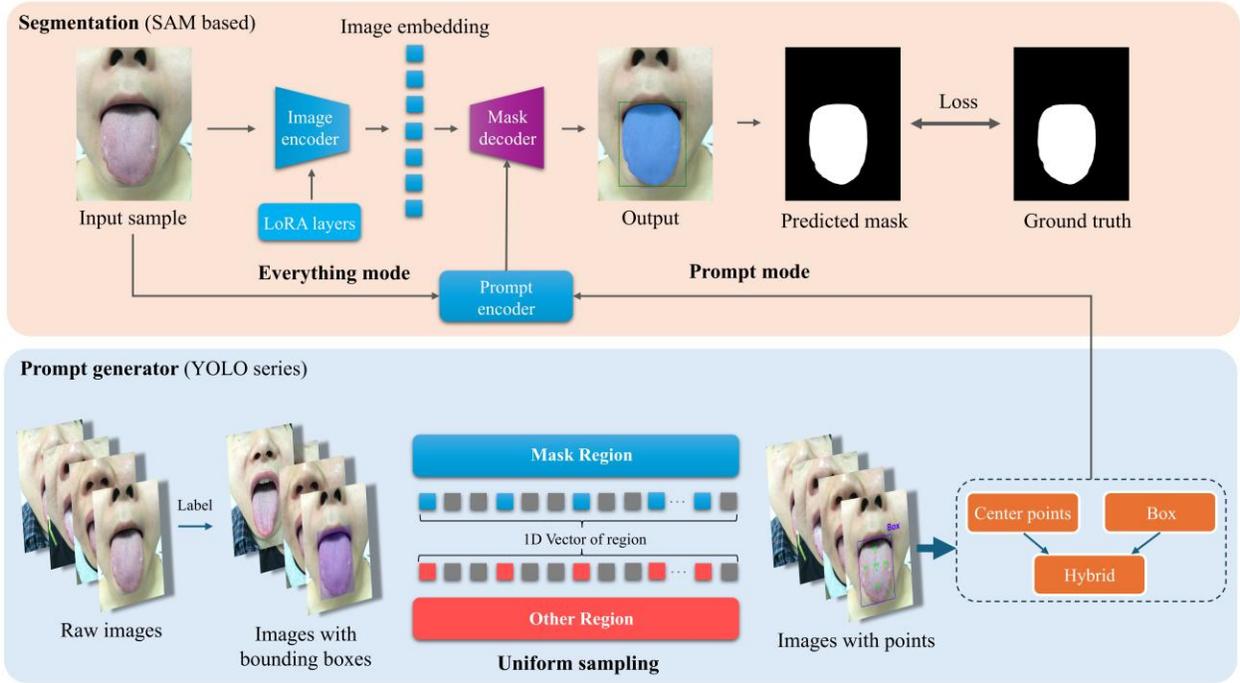

**Fig. 3. The architecture of TOM_L: a SAM-based tongue image segmentation model.** The framework consists of two components. The lower part illustrates the prompt generation process, where bounding boxes and center points are sampled from annotated tongue images to create box, point, or hybrid prompts. These prompts are fed into the upper segmentation pipeline, where the SAM backbone is enhanced with LoRA layers and partially fine-tuned. Specifically, only the prompt encoder and mask decoder of SAM are updated, while the image encoder is adapted via LoRA. The model, referred to as TOM_L, outputs predicted tongue masks supervised by ground-truth labels.

### 3.2.1. Prompt generation module

To enhance the segmentation accuracy and adaptability, we design a prompt generation framework that provides spatial priors in the form of box, point, and hybrid prompts. Given a dataset of tongue images with annotated segmentation masks, we first used a pretrained YOLOv9 (C.-Y. Wang et al., 2025) object detector to identify the primary tongue region in each image. The resulting bounding boxes are used as coarse region-level prompts and referred to as box prompts. To incorporate more detailed spatial guidance, we further sample center points from both foreground and background regions. Foreground points are randomly selected within the ground truth mask area, while background points are sampled from regions outside the mask. This balanced sampling simulates interactive segmentation scenarios and helps the model distinguish between tongue and background.

from YOLO-predicted bounding boxes; (2) point prompts, consisting of the sampled foreground and background points; and (3) hybrid prompts, which combine both boxes and points to provide comprehensive spatial context. This strategy allows TOM_L to better localize ambiguous tongue boundaries, improves generalization under varied lighting and pose conditions, and lays the foundation for interactive segmentation scenarios.

### 3.2.2. Segmentation module

To better adapt SAM to the tongue image domain, we introduce a task-specific fine-tuning strategy while retaining its efficient prompt-driven architecture. Specifically, TOM_L retains the original modular design of SAM, consisting of an image encoder, a prompt encoder, and a mask decoder. The image encoder is a vision transformer (ViT-H) responsible for extracting high-level image embeddings. We integrate LoRA layers into the image encoder while keeping the pre-trained SAM image encoder frozen during training. Only the LoRA modules are trained, enabling TOM_L to effectively adapt to tongue segmentation tasks. This approach ensures efficient adaptation to the target dataset while introducing only a minimal number of additional parameters.

Meanwhile, the prompt encoder and mask decoder are fully fine-tuned to enable effective integration of prompt signals and improve segmentation precision. The prompt encoder processes box, point, or hybrid prompts generated from the prompt generation module, encoding them into sparse and dense tokens. These are fused with image embeddings in the mask decoder to produce low-resolution masks, which are then upsampled to match the input image resolution. During training, the predicted masks are supervised by binary ground truth masks using a combination of Binary Cross-Entropy (BCE) (Guo et al., 2023) and Dice loss (Jadon, 2020). This design allows TOM_L to benefit from the generalization ability of SAM while tailoring its attention to the tongue image segmentation domain through efficient parameter adaptation and prompt-aware refinement.

### 3.3. Multi-teacher Knowledge distillation

To enhance the performance of a lightweight student model while preserving the rich semantic and structural information captured by multiple teacher models, as shown in **Fig. 4** and **Algorithm 1**, we propose a hybrid multi-teacher distillation framework. Our approach integrates logits-based

KD using Kullback-Leibler (KL) (Ji et al., 2022) divergence and mask-level distillation using mean squared error (MSE), ensuring a more comprehensive knowledge transfer from the teacher models to the student model. Specifically, we employ three teacher models—SAM (TOM_L based), Unet (ResNet34-based), and DeepLabV3 (ResNet50-based)—each trained independently to segment tongue images with high accuracy. Unlike conventional KD, which focuses solely on pixel-wise losses, our framework distills both soft and hard knowledge from teacher models, facilitating better generalization and robustness in the student model.

Given an input image $I$, each teacher model generates both logits and final segmentation masks as follows:

$$logits_{SAM}, logits_{Unet}, logits_{DeepLab} \leftarrow T_{SAM}(I), T_{Unet}(I), T_{DeepLab}(I) \quad (4)$$

$$P_{SAM}, P_{Unet}, P_{DeepLab} \leftarrow T_{SAM}(I), T_{Unet}(I), T_{DeepLab}(I) \quad (5)$$

Simultaneously, the student model $S$, based on a TinyViT (K. Wu et al., 2022) backbone, produces its own logits as:

$$logits_S \leftarrow S(I), P_S \leftarrow \sigma(logits_S) \quad (6)$$

To align the student's predictions with those of the teacher models, we introduce three complementary loss functions. The first is Logits-Based KL divergence loss. A temperature scaling factor $T$ is introduced to soften the logits and facilitate knowledge transfer as:

$$\mathcal{L}_{KL} = \alpha \cdot KL(\sigma(logits_{SAM}/T)||\sigma(logits_S/T)) + \beta \cdot KL(\sigma(logits_{Unet}/T)||\sigma(logits_S/T)) \\ + \gamma \cdot KL(\sigma(logits_{DeepLab}/T)||\sigma(logits_S/T)) \quad (7)$$

where: $KL(P||Q)$ measures the difference between the predicted probability distributions of the student and teacher models. $T$ (temperature hyperparameter) softens the probability distribution to prevent overfitting to teacher predictions, and $\alpha, \beta, \gamma$ are weight coefficients controlling the contribution of each teacher.

The second one is mask-level MSE loss, which directly minimizes the pixel-wise differences between the predicted segmentation maps of the student and teacher models as:

$$\mathcal{L}_{MSE} = \alpha \cdot \mathcal{L}_{MSE}(P_{SAM}, P_S) + \beta \cdot \mathcal{L}_{MSE}(P_{Unet}, P_S) + \gamma \cdot \mathcal{L}_{MSE}(P_{DeepLab}, P_S) \quad (8)$$

This loss ensures that the pixel-wise segmentation predictions of the student model are structurally aligned with those of the teacher models.

To further refine the segmentation quality, we incorporate BCE loss, ensuring that the student model maintains consistency with the ground truth labels:

$$\mathcal{L}_{BCE} = BCE(P_s, Y) \tag{9}$$

This loss serves as direct supervision, penalizing incorrect classifications at the pixel level.

To balance the contributions of different losses, we define the final objective function:

$$\mathcal{L}_{Total} = \lambda_{KL}\mathcal{L}_{KL} + \lambda_{MSE}\mathcal{L}_{MSE} + \lambda_{BCE}\mathcal{L}_{BCE} \tag{10}$$

where: $\lambda_{KL}, \lambda_{MSE}, \lambda_{BCE}$ control the relative importance of logits-based, mask-based, and ground truth supervision losses. Optimal values for these hyperparameters are determined through empirical tuning.
The student model is trained using the AdamW optimizer (Yao et al., 2021) with a learning rate of 0.0001. To prevent overfitting, we apply an early stopping criterion, halting training if the validation loss does not improve for 20 consecutive epochs. The model checkpoint with the lowest validation loss is selected as the final student model $S^*$.

During the testing phase, the optimized student model $S^*$ is applied to unseen tongue images $I_t$. The probabilistic segmentation is computed as:

$$\text{logits}_S = S^*(I_t), \hat{Y}_t = \sigma(\text{logits}_S) \tag{11}$$

Where $\hat{Y}_t$ represents the final predicted segmentation mask for the input image.

By integrating multiple teacher models, our framework effectively distills high-quality segmentation knowledge into a compact and efficient student model.

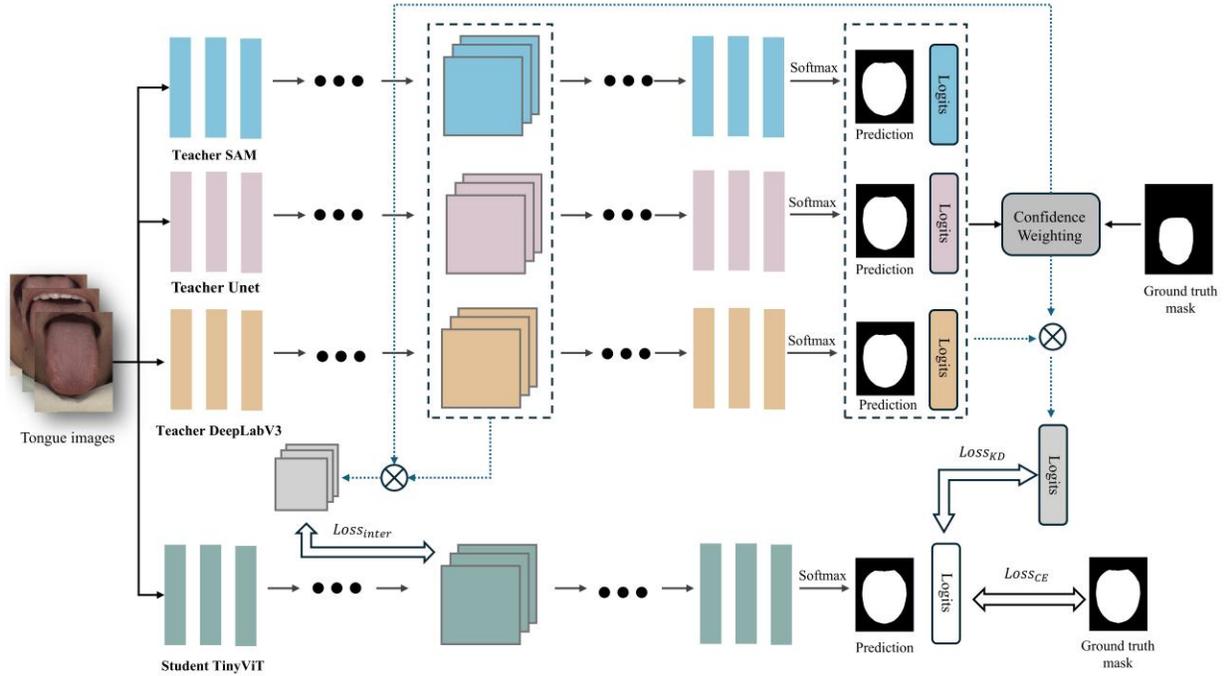

**Fig. 4. Multi-teacher distillation framework for tongue image segmentation.** The input tongue images are fed into three teacher models (Teacher SAM, Teacher UNet, and Teacher DeepLabV3) as well as a student model (Student TinyViT). Each teacher model produces feature maps and segmentation predictions, followed by a Softmax operation to obtain prediction masks and corresponding logits. The logits from all three teacher models are fused through a Confidence Weighting module to generate a more reliable supervisory signal. This fused supervision is used to compute the knowledge distillation loss (Loss_KD) with the student model's output. Additionally, intermediate features from the teacher and student models are aligned to compute an intermediate feature loss (Loss_inter). The final prediction from the student model is compared with the ground truth mask using cross-entropy loss (Loss_CE). This framework leverages the collective knowledge of multiple teacher models to enhance the segmentation performance of the lightweight student model.

### Algorithm 1: Multi-Teacher Knowledge Distillation Framework

**Training input:** Labeled training dataset $D_{train} = \{(I_i, Y_i)\}_{i=1}^{N}$, where $I_i$ represents the input image and $Y_i$ denotes the corresponding ground truth segmentation mask. Teacher models $T_1, T_2, T_3$ (e.g., SAM, UNet, and DeepLabV3) trained on the same dataset.

**Training out:** Optimized student model $S^*$ that effectively distills knowledge from multiple teacher models while maintaining supervised learning consistency.

1. Train each teacher model $T_i$ (SAM, UNet, DeepLabV3) independently until convergence.

2. Extract teacher logits from each trained teacher model:
$$\text{logits}_{T_i} = T_i(I) \forall i \in \{1,2,3\}$$
3. Compute teacher probability masks by applying the sigmoid function:
$$P_{T_i} = \sigma(\text{logits}_{T_i})$$
4. Initialize student model $S$ with a lightweight architecture (e.g., TinyViT).
5. Train student model using hybrid distillation:
    - Compute student logits:
    $$\text{logits}_S = S(I)$$
    - Compute probability mask from the student model:
    $$P_S = \sigma(\text{logits}_S)$$
    - Compute KL divergence loss between teacher and student logits:
    $$\mathcal{L}_{KL} = \sum_{i=1}^{3} \lambda_i \cdot KL\big(\sigma(\log its_{T_i}/T) || \sigma(\text{logits}_S/T)\big)$$
    - Compute mask-level MSE loss:
    $$\mathcal{L}_{MSE} = \sum_{i=1}^{3} \lambda_i \cdot MSE(P_{T_i}, P_S)$$
    - Compute binary cross-entropy loss for supervised learning:
    $$\mathcal{L}_{BCE} = BCE(P_S, Y)$$
    - Compute final loss function:
    $$\mathcal{L}_{Total} = \lambda_{KL}\mathcal{L}_{KL} + \lambda_{MSE}\mathcal{L}_{MSE} + \lambda_{BCE}\mathcal{L}_{BCE}$$
    - Update student model parameters using AdamW optimizer.
6. Monitor validation loss and apply early stopping if there is no improvement for 20 consecutive epochs.
7. Save the best student model $S^*$ based on the lowest validation loss.

**Testing input:** Unseen test images $I_t$ from the dataset $D_{test}$.

**Testing output:** Predicted segmentation mask $\hat{Y}_t$.

1. Compute logits and probability map for the student model:
$$\text{logits}_{S_t} = S^*(I_t), P_{S_t} = \sigma(\text{logits}_{S_t})$$
2. Threshold probability map to obtain binary segmentation:
$$\hat{Y}_t = 1(P_{S_t} > 0.5)$$
3. Evaluate segmentation results using mIoU and Dice Score metrics.

# 4. Experiments

In this section, we conduct a series of experiments to evaluate the performance of the TOM under different medical tongue image datasets. We compare our model with the state-of-the-art methods in the field to confirm the excellent performance of our model in terms of efficiency and accuracy.

## 4.1. Datasets

The tongue image data used in our experiments were primarily obtained from four sources: a tongue image acquisition device, the *iTongue* application, publicly available tongue image datasets on the Internet, and synthetic data generated through our proposed diffusion model-based data augmentation method. As shown in **Fig. 5**, the first portion of the data was collected using a tongue image instrument by the Shanghai University of Traditional Chinese Medicine. The tongue instrument (J. Li et al., 2022) is specifically designed for capturing tongue images under stable and consistent lighting conditions. These images were collected in a dedicated tongue image acquisition laboratory within hospitals, where controlled temperature and humidity effectively minimized the influence of external environmental factors on the collection process. Ethical approval for the data collection was obtained prior to the study. This subset contains a total of 7,561 tongue images. Dataset 2 was collected through a mobile application we previously developed, named *iTongue* (Xie et al., 2021). After users consented to the data privacy agreement, they captured tongue images using their mobile devices and uploaded them to the server. The mobile devices used were primarily various versions of iPhones and Android smartphones. Due to the diversity of capture devices and environments, the tongue images in this dataset exhibit considerable variation in lighting conditions and backgrounds; some images are overly dark or bright, and the positions and shapes of the tongues vary significantly. We selected and cropped a subset of tongue images that do not compromise user privacy for display. This dataset contains a total of 5,319 tongue images. Dataset 3 consists of publicly accessible tongue image data retrieved from the internet. From these sources, we selected a total of 5,526 images. Dataset 4 comprised synthetic tongue images generated by our diffusion-based data augmentation methods. We used various optimized prompts selected from a prompt pool, ensuring each generation produced tongue images with distinct styles and backgrounds. Using this method, we produced a total of 5,200 synthetic tongue images.

We used LabelMe (Russell et al., 2008) as a labeling tool to manually annotate the tongue surface regions and tongue positions within the aforementioned datasets and saved these annotations as ground truth. To reduce potential errors arising from manual labeling, we additionally invited two professional TCM practitioners to perform a second round of verification. Subsequently, the manually annotated dataset was randomly partitioned, with 80% used for training, 10% for validation, and the remaining 10% for testing.

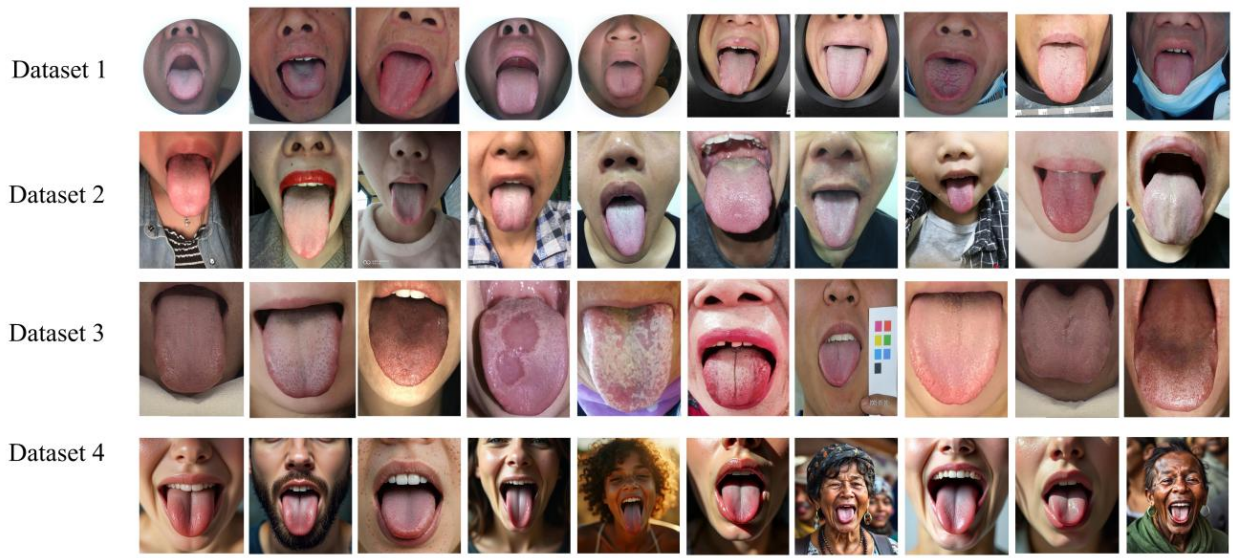

**Fig. 5. Dataset samples.** Dataset 1 consisted of tongue images collected by a tongue instrument from Shanghai University of Traditional Chinese Medicine. Dataset 2 comprised images captured with a tongue diagnosis software called *iTongue* (Xie et al., 2021). Dataset 3 contained open-access tongue images from the internet. Dataset 4 was generated using our proposed diffusion model-based data augmentation method.

### 4.2. Implementation details

All experiments are conducted using a single NVIDIA A100 GPU with 80 GB memory. The TOM_L model is initialized from the ViT-H pretrained checkpoint, with LoRA modules inserted into selected attention layers of the image encoder to reduce memory and computation overhead. During training, only the LoRA parameters, the prompt encoder, and the mask decoder are updated, while the rest of the image encoder is kept frozen. The input tongue images are resized to 1024×1024 resolution and normalized to match the SAM pretraining configuration. Ground truth masks are binarized, and bounding boxes are generated using a pretrained YOLOv9 detector. These boxes, along with uniformly sampled foreground and background points, are used to construct box, point, or hybrid prompts. The prompt encoder processes these into sparse and dense embeddings that guide the segmentation.

We use AdamW optimizer with an initial learning rate of 1e-4 and a weight decay of 1e-2. The model is trained for 300 epochs with a batch size of 64. To handle class imbalance and enhance mask quality, a compound loss function combining BCE loss and Dice loss is employed. The data augmentation methods included both traditional augmentation techniques and the diffusion model-based augmentation method we proposed earlier. All models are implemented in PyTorch, and the LoRA integration is performed using the PEFT (Parameter-Efficient Fine-Tuning) library.

In the multi-teacher distillation process, we further train two additional teacher models: a UNet model with a ResNet-34 encoder and a DeepLabV3 model with a ResNet-50 backbone. All teacher models are trained independently using the same dataset with binarized segmentation masks and we proposed data augmentation strategies. For the student, we employ TinyViT, a compact transformer-based model, equipped with a lightweight decoder composed of a 1×1 convolution and bilinear upsampling.

During distillation, the input tongue image is passed through all teacher and student models. Each teacher generates feature maps and segmentation logits, which are normalized via a Softmax layer to produce prediction masks. These logits are then aggregated using a confidence weighting module to form a fused supervision signal. The student model is trained using this fused supervision through KL divergence loss on logits. In addition, we compute an intermediate feature loss between teacher and student backbones, and a segmentation mask loss comparing the student and teacher masks. Finally, the student's prediction is compared with the ground truth using a combination of cross-entropy and Dice loss. The process is optimized with AdamW, trained for up to 300 epochs with a batch size of 64. Early stopping is applied based on validation loss with a patience of 20 epochs.

### 4.3. Evaluation of metrics

We use four evaluation methods to evaluate the accuracy of our model: mean pixel accuracy (MPA) (Möller et al., 2007), mean intersection over union (MioU) (Rezatofighi et al., 2019), Dice coefficient, and Hausdorff distance (HD) (Huttenlocher et al., 1993) are evaluated for our model.

$$MPA = \frac{1}{C}\sum_{i=1}^{C}\frac{TP_i}{TP_i + FN_i} \tag{12}$$

where C is the number of classes, and $TP_i, TN_i, FP_i, FN_i$ respectively represent the true positives, true negatives, false positives, and false negatives for class $i$.

$$MioU = \frac{1}{C}\sum_{i=1}^{C}\frac{TP_i}{TP_i + FP_i + FN_i} \tag{13}$$

Where $TP_i$ is the true positive, $FP_i$ is the false positive, and $FN_i$ is the false negative for class $i$.

$$Dice = \frac{2TP}{2TP + FP + FN} \tag{14}$$

Here, $TP$ represents the number of true positives, $FP$ the number of false positives, and $FN$ the number of false negatives.

Furthermore, the HD can be described as below:

$$H(A, B) = \max(h(A, B), h(B, A)) \tag{15}$$

Where:

$$h(A, B) = \max_{a \in A} \min_{b \in B} d(a, b) \tag{16}$$

A and B are sets of points on the predicted and true boundaries, respectively, and $d(a, b)$ is the Euclidean distance between points $a$ and $b$.

## 5. Results

### 5.1. Comparison between classic data augmentation and diffusion model-based augmentation methods

As shown in **Table 1**, we compared the performance of classic data augmentation methods with our proposed diffusion-based augmentation method. Across the evaluation metrics, the diffusion model-based augmentation consistently outperforms classic methods, exhibiting substantial improvements. For the TOM_L model, the mIoU improves from 0.9472 to 0.9765, representing a 3.09% absolute gain. The MPA increases from 0.9515 to 0.9664 (+1.56%), and the Dice coefficient shows a significant increase from 0.9731 to 0.9981 (+2.57%). Similarly, for the Unet model, the mIoU increases from 0.9121 to 0.9403 (+3.09%), MPA from 0.9370 to 0.9512 (+1.52%), and the Dice coefficient from 0.9451 to 0.9693 (+2.56%). In the case of the more advanced DeepLabV3 network, the proposed method leads to a mIoU improvement from 0.9218 to 0.9503 (+3.08%), MPA from 0.9356 to 0.9498 (+1.52%), and Dice from 0.9504 to 0.9743 (+2.51%). These consistent improvements across architectures of varying complexity validate the robustness and generality of our approach. To assess the statistical significance of the observed improvements, we performed a paired $t$-test between the performance metrics obtained using classic and diffusion-based augmentation across the three models. The resulting $p$-values for mIoU, MPA, and Dice are $p = 0.0042$, $p = 0.0068$, and $p = 0.0021$, respectively—all well below the conventional threshold of 0.05. These results indicate that the performance improvements introduced by the diffusion model-based augmentation are statistically significant.

**Table 1. Comparison between classic data augmentation methods and diffusion model-based augmentation.**

| Models | Classic methods | | | Diffusion model-based methods | | |
|---|---|---|---|---|---|---|
| | mIoU | MPA | Dice | mIoU | MPA | Dice |
| TOM_L | 0.9472 | 0.9515 | 0.9731 | 0.9765 | 0.9664 | 0.9981 |
| Unet | 0.9121 | 0.9370 | 0.9451 | 0.9403 | 0.9512 | 0.9693 |
| DeepLabV3 | 0.9218 | 0.9356 | 0.9504 | 0.9503 | 0.9498 | 0.9743 |

## 5.2. Comparison of teacher and student models in terms of architecture and parameters

**Table 2** presents the architectural configurations and parameter counts of the models involved in the knowledge distillation framework. Specifically, TOM_L, UNET, and DeepLabV3 serve as teacher models, while TOM_S is the distilled student model. The comparison highlights the differences in model complexity and computational cost between the teacher and student networks. The student model TOM_S significantly reduced parameter size compared to all three teacher models, particularly a 96.6% reduction relative to teacher model TOM_L.

**Table 2. Encoder and parameter comparison between teacher and student models**

| Models | Encoder | Parameters |
|---|---|---|
| TOM_L | ViT-H | 639M |
| UNET | ResNet 34 | 24M |
| DeepLabV3 | ResNet 50 | 43M |
| TOM_S | TinyViT | 22M |

## 5.3. Comparison of the different tongue segmentation methods

In this study, we comprehensively evaluated our proposed methods against several established segmentation approaches on the tongue image segmentation task using four distinct datasets (Dataset1, Dataset2, Dataset3, and Dataset4). The evaluation metrics included mIoU, MPA, and HD.

As illustrated in **Table 3**, our proposed methods (TOM_L and TOM_S) consistently demonstrate significant advantages across all datasets. Specifically, the TOM_L model achieved the highest mIoU and MPA values of 0.9878 and 0.9764, respectively, on Dataset 1, along with the lowest HD value of 2.186. On Dataset 2, TOM_L similarly attained optimal performance with a mIoU of 0.9719, an MPA of 0.9658, and the lowest HD of 2.158. Additionally, on Dataset 3, TOM_L maintained its superior performance, achieving the best mIoU (0.9815), MPA (0.9716), and the lowest HD (1.933) among all compared methods. Furthermore, on Dataset 4, TOM_L continued to outperform other methods, recording the highest mIoU (0.9649), highest MPA (0.9517), and a competitive HD value of 2.141.

Compared with other widely used segmentation methods such as Nested Unet, FCN, Attention Net, DeepLabV3, TongueSAM, PSPNet, SegNet, and SAM (with or without prompts), the proposed TOM models demonstrate notable improvements in both segmentation accuracy and robustness. The HD metric highlights the enhanced spatial localization accuracy of our methods, indicating that the predicted tongue boundaries are more precise and stable. In summary, the experimental results clearly validate the superior performance of our proposed TOM_L and TOM_S models for tongue image segmentation, especially in terms of accuracy (mIoU and MPA) and boundary precision (HD), significantly outperforming existing mainstream segmentation approaches.

Table 3. Comparison of tongue segmentation performance under different benchmarking methods

| Methods | Dataset1 | | | Dataset2 | | | Dataset3 | | | Dataset4 | | |
|---|---|---|---|---|---|---|---|---|---|---|---|---|
| | mIoU | MPA | HD | mIoU | MPA | DH | mIoU | MPA | HD | mIoU | MPA | HD |
| Nested Unet | 0.9361 | 0.9660 | 3.790 | 0.9415 | 0.9559 | 3.498 | 0.9574 | 0.9621 | 2.810 | 0.9261 | 0.9192 | 3.090 |
| FCN | 0.9030 | 0.9164 | 4.063 | 0.8966 | 0.9049 | 4.126 | 0.9114 | 0.9047 | 3.981 | 0.9138 | 0.9057 | 3.743 |
| Attention Net | 0.9466 | 0.9338 | 2.982 | 0.9238 | 0.9357 | 3.204 | 0.9435 | 0.9536 | 3.309 | 0.9216 | 0.9198 | 2.989 |
| DeepLabV3 | 0.9635 | 0.9613 | 2.586 | 0.9462 | 0.9538 | 2.798 | 0.9536 | 0.9574 | 2.710 | 0.9381 | 0.9267 | 2.709 |
| TongueSAM | *0.9638* | 0.9573 | *2.387* | 0.9413 | *0.9382* | 2.593 | 0.9573 | 0.9612 | 2.638 | 0.9289 | 0.9193 | 3.282 |
| PSPNet | 0.9312 | 0.9437 | 3.328 | 0.9236 | 0.9393 | 3.575 | 0.9557 | 0.9614 | 2.981 | 0.9267 | 0.9196 | 3.297 |
| SegNet | 0.9091 | 0.9164 | 3.699 | 0.9166 | 0.9049 | 3.487 | 0.9214 | 0.9147 | 3.609 | 0.9038 | 0.9157 | 3.187 |
| SAM without prompt | 0.9273 | 0.9347 | 3.356 | 0.9137 | 0.9279 | 3.508 | 0.9467 | 0.9318 | 3.108 | 0.9167 | 0.9087 | 3.536 |
| SAM with prompt | 0.9549 | 0.9612 | 2.684 | 0.9436 | *0.9572* | 2.289 | *0.9614* | 0.9534 | *2.198* | 0.9320 | *0.9413* | 3.017 |
| **TOM_L (Ours)** | **0.9878** | **0.9764** | **2.186** | **0.9719** | **0.9658** | **2.158** | **0.9815** | **0.9716** | **1.933** | **0.9649** | **0.9517** | **2.141** |
| **TOM_S (Ours)** | 0.9574 | *0.9669* | 2.549 | *0.9562* | 0.9418 | 3.259 | 0.9517 | *0.9649* | 2.571 | *0.9434* | 0.9281 | *2.662* |

The method with the highest performance on a dataset is highlighted in **bold**, while the second-best method is shown in *italics*.

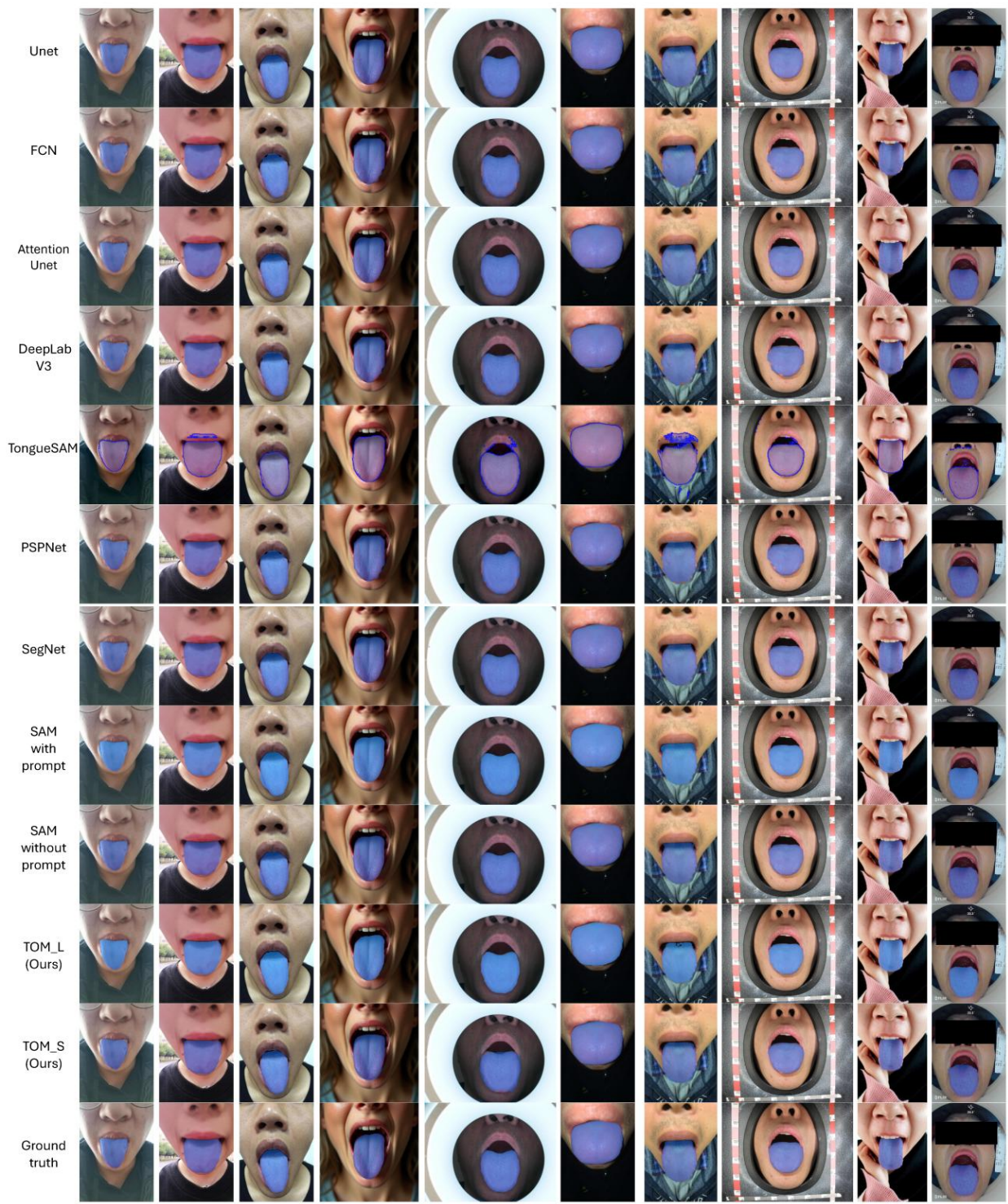

Fig. 6. **Samples of segmentation using different methods with zero-shot.** The blue area represents the segmentation output of the model. Some biometric features of the subjects in the images have been cropped or masked for privacy protection.

# 6. Application of TOM

To facilitate the use of our model by TCM practitioners to obtain tongue segment information, we have applied and deployed our trained model. Based on the TOM_S model, we have launched an online platform, as shown in **Fig. 7**, where users can bulk-upload tongue images and bulk-download the corresponding masks. For sensitive tongue surface information, we also support practitioners in obtaining local tongue masks. Users can download the local application based on TOM_L from the website and easily install the executable package to start using it. The entire installation and usage process does not require any internet connection, allowing the application to function offline and ensuring no concerns about data security. The local prediction process does not require high-end hardware configurations on the user's computer, and the local application also supports using the CPU to complete the entire tongue mask segmentation process. Whether using the online platform or the local application, users do not need to configure any development environment or possess any programming experience. With an intuitive user interface, they can easily obtain the predicted mask results through simple interactions.

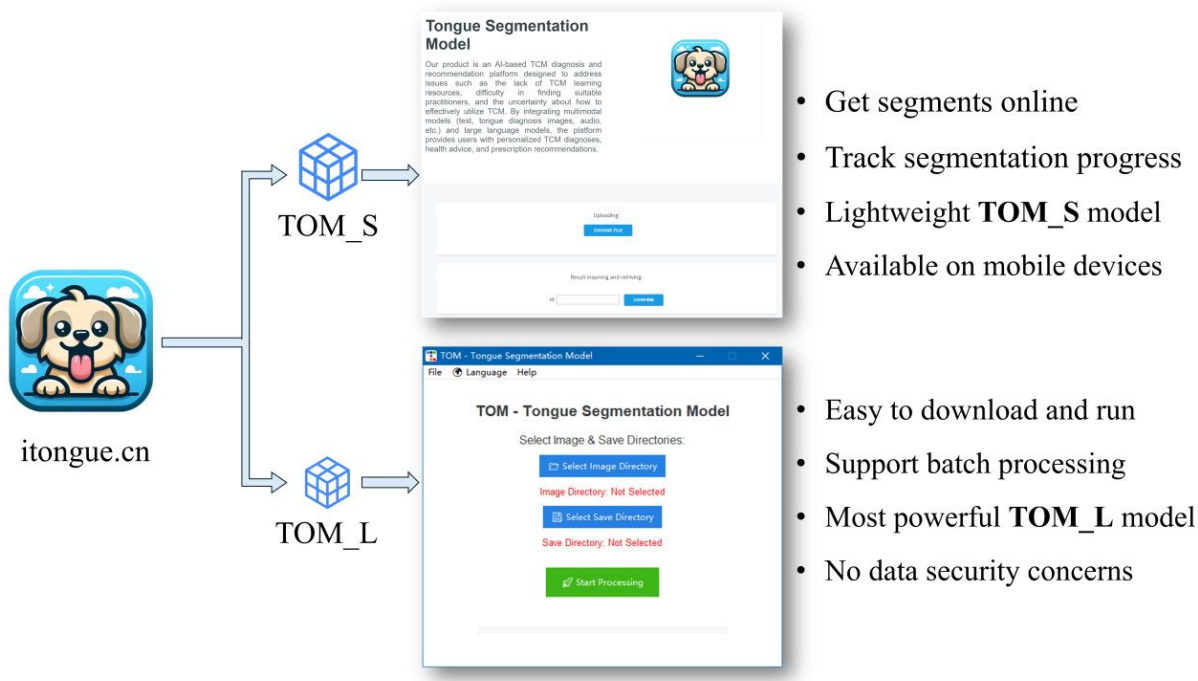

**Fig. 7. TOM local and online application.** Both models are designed for use without any programming experience. The local model, TOM_L, utilizes a teacher network to perform batch

segmentation of tongue images entirely offline, eliminating the need for an internet connection and ensuring data privacy. The lightweight online model, TOM_S, enables image prediction directly within a web browser across different operating systems, and provides real-time tracking of the prediction progress.

## 7. Case study of TCM constitution classification

The theory of constitutional types is an essential component of TCM. According to TCM principles, the constitution represents an individual's innate and acquired characteristics throughout their lifespan, serving as an integrated manifestation of physiological functions and psychological states. Different constitutional types exhibit strong correlations with certain diseases and can even predispose individuals to particular illnesses (Sun et al., 2014). Since 2005, the constitutional classification proposed by Wang Qi has been widely recognized as a standard for categorizing TCM constitutions (L. Li et al., 2021). According to the national standard published in 2009 by the China Association of Chinese Medicine, TCM constitutions can be classified into the following nine categories: Balanced Constitution (Pinghe), Qi-deficiency Constitution (Qixu), Yang-deficiency Constitution (Yangxu), Yin-deficiency Constitution (Yinxu), Phlegm-dampness Constitution (Tanshi), Damp-heat Constitution (Shire), Blood-stasis Constitution (Xueyu), Qi-stagnation Constitution (Qiyu), and Special Constitution (Tebing).

Tongue diagnosis is an effective method in TCM for identifying an individual's constitution. TCM practitioners determine constitutional types by observing features of the tongue surface, such as color, morphology, and moisture. In our previous work (Xie et al., 2021), we explored the use of ResNet50 for classifying TCM constitutions based on tongue images. Specifically, we utilized a dataset consisting of 2,215 tongue images captured by tongue imaging instruments. The dataset was partitioned into training (80%), validation (10%), and testing (10%) subsets. Following experimental evaluations, the ResNet50 architecture demonstrated the best classification performance. Under identical computational environments, we separately trained the network using the original tongue images and segmented tongue-region images. Finally, we compared and analyzed the classification performances obtained from these two sets of input data to assess the effectiveness of using segmented tongue regions for TCM constitution classification.

**Table 4** shows the performance differences when using different datasets with the same network structure. It can be observed that the classification accuracy of the model using tongue segment data improved from 68.18% with the original full-color tongue image data to 64.52%, the F1 score increased from 0.6486 to 0.8046, and the MCC improved from 0.6253 to 0.6484. Although the performance improvement is not particularly significant, there is still an overall enhancement in classification performance. We also plotted the feature maps of the classification model using the original data. Some of the feature regions in these maps sampled information from the background of the tongue image. This suggests that the classification model for body constitution was not solely based on the tongue surface information, and background noise played a role in the constitution classification, which presents challenges for the interpretability of intelligent TCM body constitution classification. However, as shown in **Fig. 8,** when the tongue segment data, which removed background information, was used as input for the constitution classification model, all the feature maps were concentrated on the tongue surface. It became evident which areas of the tongue surface contributed to the classification of the constitution, thereby enhancing the potential for the interpretability of TCM body constitution classification based on tongue surface features.

**Table 4. Performance comparison of TCM constitution prediction using original tongue images versus segmented tongue-region images.**

|  | Accuracy | MCC | F1 |
|---|---|---|---|
| Raw tongue images | 0.6452 | 0.6253 | 0.6486 |
| Tongue segments | 0.6818 | 0.6484 | 0.8046 |

Accurate segmentation of the tongue surface region provides an effective pathway for the interpretability of intelligent tongue diagnosis, enabling researchers to deeply analyze, understand, and grasp the correlation between the features of different tongue surface areas and specific diseases. This, in turn, promotes the theoretical and practical development of intelligent tongue diagnosis. Furthermore, it further reinforces the importance of tongue surface segmentation as a fundamental task in intelligent TCM tongue diagnosis, with its quality directly impacting the accuracy and reliability of subsequent tasks such as tongue image analysis and body constitution classification. This offers more opportunities for the development and research in downstream intelligent diagnosis, classification, and health assessment.

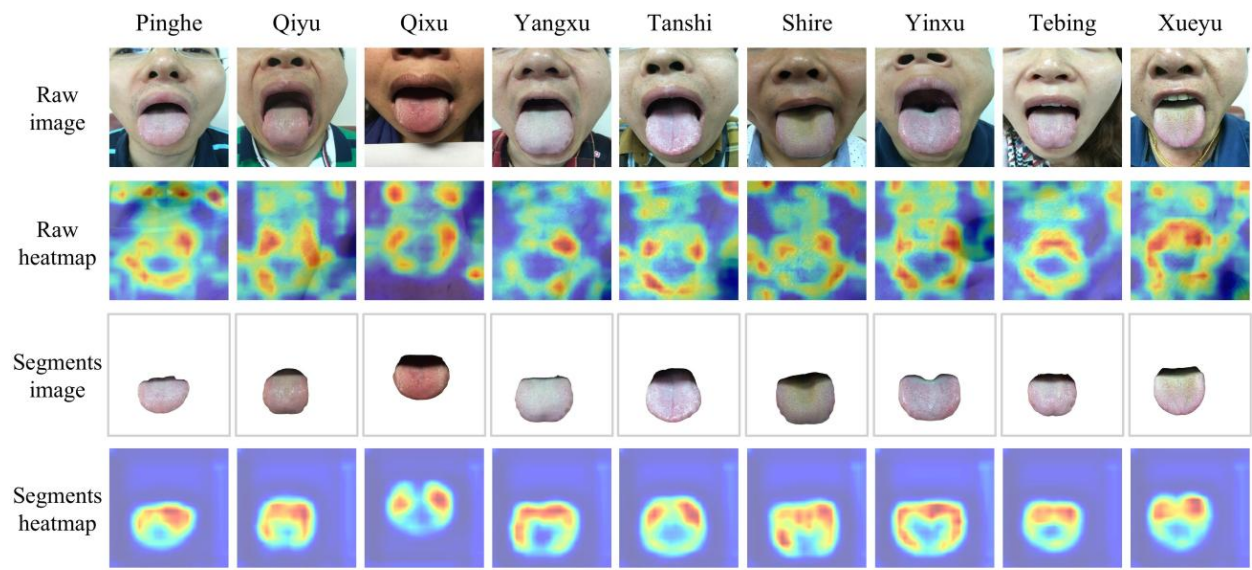

**Fig. 8. Visualization results of each body constitution type and the corresponding feature maps from ResNet50 output.** The first row shows the original tongue images. As seen in the second row, using raw images for classification may lead to the model extracting features from background areas, which are clearly not meaningful for constitution classification and reduce the interpretability of the results. However, as shown in the third and fourth rows, when using pure tongue segments as training data, all extracted features are located on the tongue surface, revealing how different regions of the tongue contribute to classification outcomes.

## 8. Discussion

Diffusion models have demonstrated strong capabilities in medical image synthesis due to their generation of high-quality, diverse data. In this study, we employed them to synthesize tongue images exhibiting variations in background, illumination, color, and texture, which were used to enhance segmentation model training. Experimental results indicated improved generalization, suggesting diffusion-based augmentation as a viable strategy for limited tongue image datasets. Nonetheless, the iterative denoising process inherent to diffusion models incurs significant computational costs, particularly for high-resolution image generation. Additionally, despite the realism of diffusion-generated tongue images, subtle inconsistencies remain compared to clinical counterparts, such as unnatural reddish tips or unrealistic fissure patterns, potentially compromising diagnostic reliability. In some cases, images generated from text prompts resemble animal, rather than human, tongues—possibly due to the scarcity of human tongue data online and the inclusion of animal tongue images in training datasets on certain platforms. As a result, manual screening of synthetic images is often necessary. Moreover, since most diffusion models are unsupervised, the majority of generated samples still require manual annotation, increasing labor demands. While multi-teacher distillation allows student models to approach teacher-level performance, it faces limitations, including high computational overhead, prediction conflicts among teachers, and performance saturation due to redundancy—potentially causing overfitting and reduced label sensitivity. The trained models were deployed on both web-based and local platforms for free access by TCM practitioners and researchers. The web tool offers cross-platform compatibility and ease of maintenance but incurs server costs and depends on network conditions, raising potential data privacy concerns. In contrast, the local tool supports offline use and enhanced data security, making it suitable for clinical settings, though it requires sufficient hardware and involves higher cross-platform development costs.

In future work, we will further optimize the tool's usability, including extending support to additional platforms such as Linux and MacOS. For synthetic data augmentation, we will explore the use of conditional diffusion models to guide the generation of synthetic data, thereby more effectively and broadly supporting a variety of downstream tasks, including tongue image classification and tongue diagnosis.

## 9. Conclusion

This paper presents a tongue image segmentation method based on multi-teacher knowledge distillation, incorporating a novel data augmentation strategy during training distinct from traditional approaches. Specifically, synthetic tongue image data are generated by diffusion models, significantly enhancing the robustness and generalization capabilities of the segmentation model. Experimental results demonstrate that the proposed approach achieves superior performance in terms of segmentation accuracy and model compactness compared to existing methods. Additionally, we present a case study illustrating that, compared with raw tongue images, segmented tongue-region data yield improved classification performance and interpretability in TCM constitution classification tasks. Finally, the proposed model is deployed in both web-based and offline tool versions, providing comprehensive support for downstream tongue segmentation and diagnostic tasks, thus promoting the rapid advancement of digitalized TCM.

## CRediT authorship contribution statement

**Jiacheng Xie:** Writing – original draft, Methodology, Data curation, Formal analysis. **Ziyang Zhang:** Software, Visualization, Methodology. **Biplab Poudel:** Validation, Methodology. **Congyu Guo:** Software, Data curation. **Yang Yu:** Validation, Writing – review and editing. **Guanghui An:** Data curation, Resources. **Xiaoting Tang:** Data curation, Resources. **Lening Zhao:** Validation. **Chunhui Xu:** Validation. **Dong Xu:** Supervision, Funding acquisition, Writing – review and editing, Project administration.

## Declaration of competing interest

The authors declare that they have no known competing financial interests or personal relationships that could have appeared to influence the work reported in this paper.

## Acknowledgements

This work is partially supported by the Paul K. and Diane Shumaker Endowment Fund to Dong Xu. The authors would like to thank the Research Computing Support Services at the University of Missouri for providing high-performance computing infrastructure.

## Data availability

The open-source datasets used in this study are available at https://itongue.cn/data_list.html. The synthetic datasets generated during this work are available upon reasonable request. Patient-related private datasets cannot be shared publicly due to privacy and ethical restrictions.

# References

Adya, K. A., Inamadar, A. C., & Palit, A. (2018). The strawberry tongue: What, how and where? *Indian Journal of Dermatology, Venereology and Leprology*, *84*, 500. https://doi.org/10.4103/ijdvl.IJDVL_57_17

Al-amri, S. S., Kalyankar, N. V., & D, K. S. (2010). *Image segmentation by using threshold techniques* (No. arXiv:1005.4020). arXiv. https://doi.org/10.48550/arXiv.1005.4020

Avraham, K. B., Schickler, M., Sapoznikov, D., Yarom, R., & Groner, Y. (1988). Down's syndrome: Abnormal neuromuscular junction in tongue of transgenic mice with elevated levels of human Cu/Zn-superoxide dismutase. *Cell*, *54*(6), 823–829. https://doi.org/10.1016/S0092-8674(88)91153-1

Azad, R., Aghdam, E. K., Rauland, A., Jia, Y., Avval, A. H., Bozorgpour, A., Karimijafarbigloo, S., Cohen, J. P., Adeli, E., & Merhof, D. (2024). Medical image segmentation review: The success of U-net. *IEEE Transactions on Pattern Analysis and Machine Intelligence*, *46*(12), 10076–10095. https://doi.org/10.1109/TPAMI.2024.3435571

Bertelli, L., Yu, T., Vu, D., & Gokturk, B. (2011). Kernelized structural SVM learning for supervised object segmentation. *CVPR 2011*, 2153–2160. https://doi.org/10.1109/CVPR.2011.5995597

*BrushNet: A plug-and-play image inpainting model with decomposed dual-branch diffusion | computer vision – ECCV 2024* (world). (n.d.). Guide Proceedings. https://doi.org/10.1007/978-3-031-72661-3_9

Buslaev, A., Iglovikov, V. I., Khvedchenya, E., Parinov, A., Druzhinin, M., & Kalinin, A. A. (2020). Albumentations: Fast and Flexible Image Augmentations. *Information*, *11*(2), Article 2. https://doi.org/10.3390/info11020125

Cao, S., Wu, Q., & Ma, L. (2023). TongueSAM: An universal tongue segmentation model based on SAM with zero-shot. *2023 IEEE International Conference on Bioinformatics and Biomedicine (BIBM)*, 4520–4526. https://doi.org/10.1109/BIBM58861.2023.10385570

Chen, J., & Bai, X. (2023). *Learning to "segment anything" in thermal infrared images through knowledge distillation with a large scale dataset SATIR* (No. arXiv:2304.07969). arXiv. https://doi.org/10.48550/arXiv.2304.07969

*Design guidelines for prompt engineering text-to-image generative models | proceedings of the 2022 CHI conference on human factors in computing systems* (world). (n.d.). ACM Conferences. https://doi.org/10.1145/3491102.3501825

Dosovitskiy, A., Beyer, L., Kolesnikov, A., Weissenborn, D., Zhai, X., Unterthiner, T., Dehghani, M., Minderer, M., Heigold, G., Gelly, S., Uszkoreit, J., & Houlsby, N. (2021). *An image is worth 16x16 words: Transformers for image recognition at scale* (No. arXiv:2010.11929). arXiv. https://doi.org/10.48550/arXiv.2010.11929

Farman, A. G. (1976). Atrophic lesions of the tongue: A prevalence study among 175 diabetic patients. *Journal of Oral Pathology & Medicine*, *5*(5), 255–264. https://doi.org/10.1111/j.1600-0714.1976.tb01774.x

Guo, Q., Wang, C., Xiao, D., & Huang, Q. (2023). A novel multi-label pest image classifier using the modified swin transformer and soft binary cross entropy loss. *Engineering Applications of Artificial Intelligence*, *126*, 107060. https://doi.org/10.1016/j.engappai.2023.107060

He, X., Zhou, Y., Zhao, J., Zhang, D., Yao, R., & Xue, Y. (2022). Swin transformer embedding UNet for remote sensing image semantic segmentation. *IEEE Transactions on Geoscience and Remote Sensing*, *60*, 1–15. https://doi.org/10.1109/TGRS.2022.3144165

Hinton, G., Vinyals, O., & Dean, J. (2015). *Distilling the knowledge in a neural network* (No. arXiv:1503.02531). arXiv. https://doi.org/10.48550/arXiv.1503.02531

Hsiao, W.-L., Katsman, I., Wu, C.-Y., Parikh, D., & Grauman, K. (2019). *Fashion++: Minimal edits for outfit improvement*. 5047–5056. https://openaccess.thecvf.com/content_ICCV_2019/html/Hsiao_Fashion_Minimal_Edits_for_Outfit_Improvement_ICCV_2019_paper.html

Hu, M., Li, Y., & Yang, X. (2023). *SkinSAM: Empowering skin cancer segmentation with segment anything model* (No. arXiv:2304.13973). arXiv. https://doi.org/10.48550/arXiv.2304.13973

Huang, K., Zhou, T., Fu, H., Zhang, Y., Zhou, Y., Gong, C., & Liang, D. (2025). Learnable prompting SAM-induced knowledge distillation for semi-supervised medical image segmentation. *IEEE Transactions on Medical Imaging*, 1–1. https://doi.org/10.1109/TMI.2025.3530097

Huttenlocher, D. P., Klanderman, G. A., & Rucklidge, W. J. (1993). Comparing images using the hausdorff distance. *IEEE Transactions on Pattern Analysis and Machine Intelligence*, *15*(9), 850–863. https://doi.org/10.1109/34.232073

Ibragimov, B., Prince, J. L., Murano, E. Z., Woo, J., Stone, M., Likar, B., Pernuš, F., & Vrtovec, T. (2015). Segmentation of tongue muscles from super-resolution magnetic resonance images. *Medical Image Analysis*, *20*(1), 198–207. https://doi.org/10.1016/j.media.2014.11.006

Jadon, S. (2020). A survey of loss functions for semantic segmentation. *2020 IEEE Conference on Computational Intelligence in Bioinformatics and Computational Biology (CIBCB)*, 1–7. https://doi.org/10.1109/CIBCB48159.2020.9277638

Ji, S., Zhang, Z., Ying, S., Wang, L., Zhao, X., & Gao, Y. (2022). Kullback–leibler divergence metric learning. *IEEE Transactions on Cybernetics*, *52*(4), 2047–2058. https://doi.org/10.1109/TCYB.2020.3008248

Jiang, F., Xu, Z., Niu, L., Xiang, Z., Ramasubramanian, B., Li, B., & Poovendran, R. (2024). ArtPrompt: ASCII art-based jailbreak attacks against aligned LLMs. In L.-W. Ku, A. Martins, & V. Srikumar (Eds.), *Proceedings of the 62nd Annual Meeting of the Association for Computational Linguistics (Volume 1: Long Papers)* (pp. 15157–15173). Association for Computational Linguistics. https://doi.org/10.18653/v1/2024.acl-long.809

Julka, S., & Granitzer, M. (2024). Knowledge distillation with segment anything (SAM) model for planetary geological mapping. In G. Nicosia, V. Ojha, E. La Malfa, G. La Malfa, P. M. Pardalos, & R. Umeton (Eds.), *Machine Learning, Optimization, and Data Science* (pp. 68–77). Springer Nature Switzerland. https://doi.org/10.1007/978-3-031-53969-5_6

Kirillov, A., Mintun, E., Ravi, N., Mao, H., Rolland, C., Gustafson, L., Xiao, T., Whitehead, S., Berg, A. C., Lo, W.-Y., Dollar, P., & Girshick, R. (2023). *Segment anything*. 4015–4026. https://openaccess.thecvf.com/content/ICCV2023/html/Kirillov_Segment_Anything_ICCV_2023_paper.html

Korshunova, I., Shi, W., Dambre, J., & Theis, L. (2017). *Fast face-swap using convolutional neural networks*. 3677–3685. https://openaccess.thecvf.com/content_iccv_2017/html/Korshunova_Fast_Face-Swap_Using_ICCV_2017_paper.html

Li, J., Huang, J., Jiang, T., Tu, L., Cui, L., Cui, J., Ma, X., Yao, X., Shi, Y., Wang, S., Wang, Y., Liu, J., Li, Y., Zhou, C., Hu, X., & Xu, J. (2022). A multi-step approach for tongue image classification in patients with diabetes. *Computers in Biology and Medicine*, *149*, 105935. https://doi.org/10.1016/j.compbiomed.2022.105935

Li, J., Zhu, H., Chen, T., & Qian, X. (2023). Generalizable pancreas segmentation via a dual self-supervised learning framework. *IEEE Journal of Biomedical and Health Informatics*, *27*(10), 4780–4791. https://doi.org/10.1109/JBHI.2023.3294278

Li, L., Wang, Z., Wang, J., Zheng, Y., Li, Y., & Wang, Q. (2021). Enlightenment about using TCM constitutions for individualized medicine and construction of Chinese-style precision medicine: Research progress with TCM constitutions. *Science China Life Sciences*, *64*(12), 2092–2099. https://doi.org/10.1007/s11427-020-1872-7

Lin, B., Xie, J., Li, C., & Qu, Y. (2018). Deeptongue: Tongue Segmentation Via Resnet. *2018 IEEE International Conference on Acoustics, Speech and Signal Processing (ICASSP)*, 1035–1039. https://doi.org/10.1109/ICASSP.2018.8462650

Ma, J., He, Y., Li, F., Han, L., You, C., & Wang, B. (2024). Segment anything in medical images. *Nature Communications*, *15*(1), Article 1. https://doi.org/10.1038/s41467-024-44824-z

Mahapatra, D. (2014). Analyzing training information from random forests for improved image segmentation. *IEEE Transactions on Image Processing*, *23*(4), 1504–1512. https://doi.org/10.1109/TIP.2014.2305073

Möller, M., Lymburner, L., & Volk, M. (2007). The comparison index: A tool for assessing the accuracy of image segmentation. *International Journal of Applied Earth Observation and Geoinformation*, *9*(3), 311–321. https://doi.org/10.1016/j.jag.2006.10.002

Núñez Amin Dick, T., Rocha Santos, L., Carneiro, S., Moore, D., Pestana, S., Laerte Boechat, J., & Lavinas Sayed Picciani, B. (2021). Investigation of oral atopic diseases: Correlation between geographic tongue and fungiform papillary glossitis. *Journal of Stomatology,*

*Oral and Maxillofacial Surgery*, *122*(3), 283–288.

https://doi.org/10.1016/j.jormas.2020.05.025

Ozgursoy, O. B., Karatayli Ozgursoy, S., Tulunay, O., Kemal, O., Akyol, A., & Dursun, G. (2009). Melkersson-Rosenthal syndrome revisited as a misdiagnosed disease. *American Journal of Otolaryngology*, *30*(1), 33–37. https://doi.org/10.1016/j.amjoto.2008.02.004

Qin, D., Bu, J.-J., Liu, Z., Shen, X., Zhou, S., Gu, J.-J., Wang, Z.-H., Wu, L., & Dai, H.-F. (2021). Efficient medical image segmentation based on knowledge distillation. *IEEE Transactions on Medical Imaging*, *40*(12), 3820–3831. https://doi.org/10.1109/TMI.2021.3098703

Ren, J., Zheng, Y., Du, H., Wang, S., Liu, L., Duan, W., Zhang, Z., Heng, L., & Yang, Q. (2020). Antibiotic-induced black hairy tongue: Two case reports and a review of the literature. *Journal of International Medical Research*, *48*(10), 0300060520961279. https://doi.org/10.1177/0300060520961279

Rezatofighi, H., Tsoi, N., Gwak, J., Sadeghian, A., Reid, I., & Savarese, S. (2019). *Generalized intersection over union: A metric and a loss for bounding box regression*. 658–666. https://openaccess.thecvf.com/content_CVPR_2019/html/Rezatofighi_Generalized_Intersection_Over_Union_A_Metric_and_a_Loss_for_CVPR_2019_paper.html

Russell, B. C., Torralba, A., Murphy, K. P., & Freeman, W. T. (2008). LabelMe: A Database and Web-Based Tool for Image Annotation. *International Journal of Computer Vision*, *77*(1), 157–173. https://doi.org/10.1007/s11263-007-0090-8

*SAM-path: A segment anything model for semantic segmentation in digital pathology | SpringerLink*. (n.d.). Retrieved March 23, 2025, from https://link.springer.com/chapter/10.1007/978-3-031-47401-9_16

Seow, J. W., Lim, M. K., Phan, R. C. W., & Liu, J. K. (2022). A comprehensive overview of deepfake: Generation, detection, datasets, and opportunities. *Neurocomputing*, *513*, 351–371. https://doi.org/10.1016/j.neucom.2022.09.135

Shi, J., Dang, J., Cui, M., Zuo, R., Shimizu, K., Tsunoda, A., & Suzuki, Y. (2021). Improvement of damage segmentation based on pixel-level data balance using VGG-unet. *Applied Sciences*, *11*(2), 518. https://doi.org/10.3390/app11020518

Shu, H., Li, W., Tang, Y., Zhang, Y., Chen, Y., Li, H., Wang, Y., & Chen, X. (2025). *TinySAM: Pushing the envelope for efficient segment anything model* (No. arXiv:2312.13789). arXiv. https://doi.org/10.48550/arXiv.2312.13789

Sun, Y., Liu, P., Zhao, Y., Jia, L., He, Y., Xue, S. A., Zheng, X., Wang, Z., Wang, N., & Chen, J. (2014). Characteristics of TCM constitutions of adult Chinese women in hong kong and identification of related influencing factors: A cross-sectional survey. *Journal of Translational Medicine*, *12*(1), 140. https://doi.org/10.1186/1479-5876-12-140

Tan, D., Zang, H., Zhang, X., Gao, H., Wang, J., Wang, Z., Zhai, X., Li, H., Tang, Y., & Han, A. (2025). Tongue-LiteSAM: A lightweight model for tongue image segmentation with zero-shot. *IEEE Access*, *13*, 11689–11703. https://doi.org/10.1109/ACCESS.2025.3528658

Tang, J.-L., Liu, B.-Y., & Ma, K.-W. (2008). Traditional Chinese medicine. *The Lancet*, *372*(9654), 1938–1940. https://doi.org/10.1016/S0140-6736(08)61354-9

*The Association of Tongue Scalloping With Obstructive Sleep Apnea and Related Sleep Pathology—Todd M. Weiss, Strahil Atanasov, Karen H. Calhoun, 2005*. (n.d.). Retrieved June 7, 2024, from https://journals.sagepub.com/doi/abs/10.1016/j.otohns.2005.07.018

Tomooka, K., Tanigawa, T., Sakurai, S., Maruyama, K., Eguchi, E., Nishioka, S., Miyoshi, N., Kakuto, H., Shimizu, G., Yamaoka, D., & Saito, I. (2017). Scalloped tongue is associated with nocturnal intermittent hypoxia among community-dwelling Japanese: The Toon Health Study. *Journal of Oral Rehabilitation*, *44*(8), 602–609. https://doi.org/10.1111/joor.12526

Tremeau, A., & Borel, N. (1997). A region growing and merging algorithm to color segmentation. *Pattern Recognition*, *30*(7), 1191–1203. https://doi.org/10.1016/S0031-3203(96)00147-1

Wang, C.-Y., Yeh, I.-H., & Mark Liao, H.-Y. (2025). YOLOv9: Learning what you want to learn using programmable gradient information. In A. Leonardis, E. Ricci, S. Roth, O. Russakovsky, T. Sattler, & G. Varol (Eds.), *Computer Vision – ECCV 2024* (pp. 1–21). Springer Nature Switzerland. https://doi.org/10.1007/978-3-031-72751-1_1

Wang, J., Wang, K., Yu, Y., Lu, Y., Xiao, W., Sun, Z., Liu, F., Zou, Z., Gao, Y., Yang, L., Zhou, H.-Y., Miao, H., Zhao, W., Huang, L., Zeng, L., Guo, R., Chong, I., Deng, B., Cheng, L., … Qu, J. (2025). Self-improving generative foundation model for synthetic medical image generation and clinical applications. *Nature Medicine*, *31*(2), 609–617. https://doi.org/10.1038/s41591-024-03359-y

Wu, J., Ji, W., Fu, H., Xu, M., Jin, Y., & Xu, Y. (2024). MedSegDiff-V2: Diffusion-based medical image segmentation with transformer. *Proceedings of the AAAI Conference on Artificial Intelligence*, *38*(6), 6030–6038. https://doi.org/10.1609/aaai.v38i6.28418

Wu, K., Zhang, J., Peng, H., Liu, M., Xiao, B., Fu, J., & Yuan, L. (2022). TinyViT: Fast pretraining distillation for small vision transformers. In S. Avidan, G. Brostow, M. Cissé, G. M. Farinella, & T. Hassner (Eds.), *Computer Vision – ECCV 2022* (pp. 68–85). Springer Nature Switzerland. https://doi.org/10.1007/978-3-031-19803-8_5

Xie, J., Jing, C., Zhang, Z., Xu, J., Duan, Y., & Xu, D. (2021). Digital tongue image analyses for health assessment. *Medical Review*, *1*(2), 172–198. https://doi.org/10.1515/mr-2021-0018

Xue, X., Lu, Z., Huang, D., Wang, Z., Ouyang, W., & Bai, L. (2024). *ComfyBench: Benchmarking LLM-based agents in ComfyUI for autonomously designing collaborative AI systems* (No. arXiv:2409.01392). arXiv. https://doi.org/10.48550/arXiv.2409.01392

Yao, Z., Gholami, A., Shen, S., Mustafa, M., Keutzer, K., & Mahoney, M. (2021). ADAHESSIAN: An adaptive second order optimizer for machine learning. *Proceedings of the AAAI Conference on Artificial Intelligence*, *35*(12), 10665–10673. https://doi.org/10.1609/aaai.v35i12.17275

Zhang, C., Han, D., Qiao, Y., Kim, J. U., Bae, S.-H., Lee, S., & Hong, C. S. (2023). *Faster segment anything: Towards lightweight SAM for mobile applications* (No. arXiv:2306.14289). arXiv. https://doi.org/10.48550/arXiv.2306.14289

Zhang, H., Zuo, W., Wang, K., & Zhang, D. (2006). A snake-based approach to automated segmentation of tongue image using polar edge detector. *International Journal of Imaging Systems and Technology*, *16*(4), 103–112. https://doi.org/10.1002/ima.20075

Zhang, K., & Liu, D. (2023). *Customized segment anything model for medical image segmentation* (No. arXiv:2304.13785). arXiv. https://doi.org/10.48550/arXiv.2304.13785

Zhong, F., Qin, C., Feng, Y., Zeng, J., Jia, X., Zhong, F., Luo, J., & Yang, M. (2025). MEAT-SAM: More efficient automated tongue segmentation model. *IEEE Access*, *13*, 5175–5192. https://doi.org/10.1109/ACCESS.2024.3522961

Zuo, W., Wang, K., Zhang, D., & Zhang, H. (2004). Combination of polar edge detection and active contour model for automated tongue segmentation. *Third International Conference on Image and Graphics (ICIG'04)*, 270–273. https://doi.org/10.1109/ICIG.2004.48